\documentclass[[preprint,nofootinbib,aps,superscriptaddress,eqsecnum,a4paper,12pt]{revtex4-2}
\usepackage[utf8]{inputenc}
\usepackage{graphicx}
\usepackage{amsmath}
\usepackage{amssymb}

\usepackage{color}
\usepackage{hyperref}
\begin{document}

	\title{Neutrinoless double-beta decay and lepton flavor violation in discrete flavor symmetric left-right symmetric model}
	\author{Bichitra Bijay Boruah}
	\email{bijay@tezu.ernet.in}
	\author{Mrinal Kumar Das}
	\email{mkdas@tezu.ernet.in}
	\affiliation{Department of Physics, Tezpur University, Tezpur 784028, India}
	
		\begin{abstract}
		\textbf{ABSTRACT:}  In this work, we have studied neutrinoless double beta decay (NDBD) and charged lepton flavor violation(CLFV) in a generic Left-Right symmetric model (LRSM). In this framework, type-I and type-II seesaw terms arise naturally. We have used $A_{4}\times Z_{2}$ discrete flavor symmetry to realize the LRSM. Within the model, we have considered type-I and type-II dominant cases and analyzed the new physics contributions to the NDBD process coming from different particles of LRSM. We tried to find the leading order contributions to NDBD in type-I and type-II seesaw along with the decay rate of the process in our work. We have also studied different charged lepton flavor violating processes such as $\mu \rightarrow 3e$ and $\mu \rightarrow e \gamma$ and correlated with neutrino mass within the model.
	\end{abstract} 
	\maketitle

	\section{INTRODUCTION}
	
	In the field of particle physics, The standard model (SM) is one of the most well-known and accepted theories to date. Although, there are quite a lot that remains unexplained within the framework of SM, such as the smallness of neutrino mass, dark matter, baryon asymmetry of the universe(BAU). The experimental evidence for neutrino oscillations coming from the atmospheric, solar, reactor, and long-baseline neutrino experiments like
	MINOS \cite{Adamson:2020jvo}, T2K \cite{Abe:2011sj}, Double Chooz \cite{Abe:2011fz}, Daya Bay \cite{Abe:2013xua}, RENO \cite{Ahn:2012nd} etc, established the existence 
	of neutrino mass and large mixing parameters. These experiments have not only confirmed the earlier observations but also measured the neutrino parameters more accurately.
	Understanding the smallness of neutrino mass and its origin has been one of the most sought-after topics amongst the scientific research community, in present times. It appeals for a better explanation of the process beyond the most successful but inadequate Standard Model (SM) of particle physics. We know that the absolute neutrino mass scale is still unperceived. However, the
	Planck experiment has given an upper bound on the sum of the light neutrino mass to be $\sum_{i} |m_{i}| <0.23$eV \cite{Planck:2013pxb}  in 2012, and recently the bound has been constrained to $\sum_{i} |m_{i}| <0.11$ eV \cite{Planck:2018vyg}. Simply the neutrino mass can be explained by introducing at least
	 two right-handed (RH) neutrino in the Standard Model (SM) which allows Dirac
	 coupling with the Higgs, like other fermions in the SM. However, corresponding Yukawa
	 coupling has to be fine-tuned to a quite unnatural extent. This kind of fine-tuning can be avoided to explain the neutrino masses in the seesaw mechanism, a mechanism
	 beyond SM (BSM) physics which is categorized into type-I \cite{Minkowski:1977sc,Yanagida:1979as,Gell-Mann:133618,Mohapatra:1979ia,Schechter:1980gr}, type-II \cite{Mohapatra:1980yp,Wetterich:1981bx,Schechter:1981cv,Brahmachari:1997cq,Antusch:2004xy}, type-III \cite{Foot:1988aq}, inverse seesaw mechanism \cite{Haba:2016lxc} etc. The BSM physics also can be useful in explaining various phenomena
	 like Baryon Asymmetry of the Universe (BAU) \cite{Borah:2014bda}, Lepton Number Violation (LNV) \cite{Bilenky:2012qi}, Lepton
	 Flavour Violation (LFV) \cite{Altarelli:2012bn,Barry:2013xxa}, the existence of dark matter \cite{Hambye:2009pw}. Present status and future prospective of neutrino physics in theoretical and experimental point of view can be found in many of the important literature such as \cite{Athar:2021xsd,Nath:2018ywc,Choubey:2021bln,Senjanovic:2020rcq}. One of the theoretical frameworks to
	 make the first three processes observable is the left-right symmetric model (LRSM), which is considered to be an appealing candidate for physics BSM. Here, the gauge
	 group is a simple extension of the SM gauge group. It provides a natural framework to
	 understand the spontaneous breaking of parity and origin of small neutrino mass via the seesaw
	 mechanism. LRSM is widely used and is an interesting theory where the left and right chiralities are treated on equal footing at high energy scales. Herein, the type-I and type-II seesaw mechanisms arise naturally. The minimal LRSM is based on the gauge group 
	 $ SU(3)_c\times SU(2)_L\times SU(2)_R\times U(1)_{B-L}$, which extends the SM to restore parity as an exact symmetry which is then broken at some intermediate-mass scale. The right-handed (RH) neutrinos are a necessary part of LRSM  which acquires a Majorana mass as soon as $SU(2)_R$ symmetry is broken at a high energy scale.

	  The nature of the neutrinos and absolute neutrino mass scale is still unknown, whether they are four-component Dirac particles possessing a conserved lepton number or two-component Majorana particles. This is directly related to the issue of LN conservation. One such process of fundamental importance in particle physics which arises in any BSM framework is neutrinoless double beta decay(NDBD/0$\nu$$\beta$$\beta$) \cite{Bilenky:2012qi, Ostrovskiy:2016uyx}. It is a second-order, and slow radioactive process that transforms a nuclide
	 of atomic number Z into its isobar with atomic number Z+2,
	 \begin{center}
	 	\begin{equation}
	 	 N(A,Z) \longrightarrow N(A,Z+2)+e^{-}+e^{-}
	 	\end{equation}
	 \end{center}
	 which violates the lepton number(LN) conservation. The main aim for the search of 0$\nu$$\beta$$\beta$ decay is the measurement of the effective Majorana neutrino mass, which is a combination of the neutrino mass eigenstates and neutrino mixing matrix terms.
	
	 Still, there is no convincing experimental evidence of the decay that exists
	 to date. But the new generation of experiments are already running or about to run to explore effective neutrino mass along with decay rates NDBD process. In addition, from the lifetime of NDBD combined with sufficient knowledge of the nuclear matrix elements (NME), one can set a constrain involving the neutrino masses. The experiments that have improved the lower bound of the half-life of the decay process include KamLANDZen \cite{KamLAND-Zen:2016pfg} and GERDA \cite{gerda} which use Xenon-136 and Germanium-76 nuclei respectively. Incorporating the results from the first and second phase of the experiment, KamLAND-Zen imposes the best lower limit on the decay half-life using Xe-136 as $T_{1/2}^{0\nu}$$>$ $1.07 \times 10^{26}$yr at $90$ percent CL and the corresponding upper limit of effective Majorana mass in the range $(0.061-0.165)$eV.  In LRSM, several new contributions appear due to the additional RH current interactions, which could lead to sizeable LFV rates for TeV scale RH neutrino that occur at rates observable in current experiments. It is found that that process $\mu\rightarrow 3e$ induced by doubly charged bosons $\Delta_{L}^{++}$ and  $\Delta_{R}^{++}$ and $\mu\rightarrow e\gamma$ provides the most significant constraint.
	 In SM the decay rates of these LFV decays are suppressed by the tiny neutrino mass, which are well below the current experimental limits and near future sensitivity.
	 No experiment so far has observed a flavor violating process involving charged leptons. However, many experiments are currently going on to set strong limits on the most relevant LFV observables that will constrain parameter space of many new physics models. The most stringent bounds on LFV come from the MEG experiment \cite{TheMEG}. The limit on branching ratio for the decay of $\mu \rightarrow e\gamma$ from this experiment is obtained to be Br($\mu \rightarrow e\gamma$)$< 4.2\times 10^{-13}$. In the case of $l_{\alpha}\rightarrow 3 l_{\beta}$ decay constrain comes from the SINDRUM experiment \cite{Perrevoort:2018cqi} is set to be $\rm BR(l_{\alpha} \rightarrow 3l_{\beta})<10^{-12}$.

	 In this work, we realized the LRSM through $A_{4}\times Z_{2}$ discrete flavor symmetry for both type-I and type-II dominant cases. Discrete symmetries constraints the Yukawa couplings of a particular model. Here, we produce a realistic neutrino mixing to do an extensive analysis of lepton number violating processes and lepton flavor violating processes. Considering different lepton flavor violating(LFV) processes such as $l_{\alpha}\rightarrow l_{\beta}\gamma$ and $l_{\alpha}\rightarrow 3l_{\beta}$($l_{\alpha}$ and $l_{\beta}$ corresponds to $\mu$ and e respectively), we analysed their impact on the neutrino phenomenology as well. Neutrinoless double beta decay($0\nu\beta\beta$) is also studied within the model by the consideration of the constraints from the KamLAND-Zen experiment \cite{KamLAND-Zen:2016pfg} and future sensitivity from LEGEND-1K experiment \cite{LEGEND:2017cdu}. Discrete groups like $ A_{4} $, $ S_{4} $ etc are used in the context of LRSM to study different phenomenology in very few works \cite{Rodejohann:2015hka, Bazzocchi:2007au, Bonilla:2020hct}

    This paper is structured as follows. In Section 2 we briefly discuss the left-right symmetric model framework and the origin of neutrino mass. Flavor symmetric models considering both type-I and type-II dominant cases are discussed in Section 3. We also discuss different new physics contributions to the amplitude of the NDBD process along with the decay rate of the process in section 4.
	 In Section 5, we briefly discussed different LFV processes and in section 6, we present our numerical analysis and results, and then in Section 7,
	we conclude by giving a brief overview of our work.

 \section{LEFT RIGHT SYMMETRIC MODEL(LRSM) AND NEUTRINO MASS}{\label{sec:level3}}

 Several groups has studied Left-right symmetric model (LRSM) since 1970's by \cite{Mohapatra:1974gc,Senjanovic:1975rk,Mohapatra:1980qe,Pati:1974yy,PhysRevD.20.776}. LRSM is a very simple extension of the standard model gauge group where parity restoration is obtained at a high energy scale and the fermions are assigned to the gauge group  $ SU(3)_c\times SU(2)_L\times SU(2)_R\times U(1)_{B-L}$ which can be tested in present-day experiments.
 The usual type-I and-II seesaw arises naturally in LRSM. Several other problems like parity violation of weak interaction, massless neutrinos, CP problems, hierarchy problems, etc can also be addressed in the framework of LRSM. The seesaw scale is identified as the breaking of the $SU(2)_R$ symmetry.
In this model, the electric charge generator is given by, $ Q=T_{3L}+T_{3R}+\frac{B-L}{2}$ \cite{Mohapatra:1979ia}, where $ T_{3L}$ and $ T_{3R}$ are the generators of $ SU(2)_L$ and $ SU(2)_R$ and B-L being the baryon minus lepton number charge operator.

The Quarks and leptons (LH and RH) that transform in the Left-Right symmetric gauge group are given by,

\begin{equation}\label{eqx}
Q_L=\left[\begin{array}{cc}
u\\
d
\end{array}\right]_L,Q_R=\left[\begin{array}{cc}
u\\
d
\end{array}\right]_R, \Psi_L=\left[\begin{array}{cc}
\nu_l\\
l
\end{array}\right]_L,\Psi_R=\left[\begin{array}{cc}
\nu_l\\
l
\end{array}\right]_R.
\end{equation}

where the Quarks are assigned with quantum numbers $(3,2,1,1/3)$ and $(3,1,2,1/3)$ and leptons with $(1,2,1,-1)$ and $(1,1,2,-1)$ respectively under
$SU(3)_c\times SU(2)_L\times SU(2)_R\times U(1)_{B-L}$. The Higgs sector in LRSM consists of a bi-doublet with quantum number $ \phi(1,2,2,0)$
and the $SU(2)_{L,R}$ triplets, $\Delta_L(1,2,1,-1)$, $\Delta_R(1,1,2,-1)$. The matrix representations are given by,

\begin{equation}\label{eqx3}
\phi=\left[\begin{array}{cc}
\phi_1^0 & \phi_1^+\\
\phi_2^- & \phi_2^0
\end{array}\right]\equiv \left( \phi_1,\widetilde{\phi_2}\right), \Delta_{L,R}=\left[\begin{array}{cc}
{\delta_\frac{L,R}{\sqrt{2}}}^+ & \delta_{L,R}^{++}\\
\delta_{L,R}^0 & -{\delta_\frac{L,R}{\sqrt{2}}}^+ .
\end{array}\right].
\end{equation}

The spontaneous symmetry breaking occurs in two successive steps given by,
$ SU(2)_L\times SU(2)_R\times U(1)_{B-L}\xrightarrow{<\Delta_R>} SU(2)_L\times U(1)_Y \xrightarrow{<\phi>} U(1)_{em}$. 
The vacuum expectation values(vev) of the neutral component of the Higgs field are $ v_R,v_L,k_1,k_2$ respectively. The vev $v_{R}$ of $\Delta_{R}$ breaks the $ SU(2)_R$ symmetry and sets the mass scale for the extra gauge bosons $ (W_R$ and Z$  \ensuremath{'})$ and right-handed neutrino
field. The vev's $ k_1$ and $ k_2$ serve the twin purpose of breaking the remaining the  $  SU(2)_L\times U(1)_{B-L}$ symmetry down to $ U(1)_{em}$, thereby setting 
the mass scales for the observed $ W_L$ and Z bosons and providing Dirac masses for the quarks and leptons. Clearly, $ v_R$ must be significantly larger than $ k_1$ and $ k_2$ for $ W_R$ and Z $\ensuremath{'}$ to have greater masses than the $W_L$ and Z bosons. $v_{L}$ is the VEV of $\Delta_L$, it plays a significant role in the seesaw relation which is the characteristics of the LR model and can be written as,

\begin{equation}\label{eqx7}
<\Delta_L>=v_{L}=\frac{\gamma k^2}{v_{R}}.
\end{equation}

\par The Yukawa lagrangian in the lepton sector is given by,
\begin{equation}\label{eq}
\mathcal{L}=h_{ij}\overline{\Psi}_{L,i}\phi\Psi_{R,j}+\widetilde{h_{ij}}\overline{\Psi}_{L,i}\widetilde{\phi}\Psi_{R,j}+f_{L,ij}{\Psi_{L,i}}^TCi\sigma_2\Delta_L\Psi_{L,j}+f_{R,ij}{\Psi_{R,i}}^TCi\sigma_2\Delta_R\Psi_{R,j}+h.c.
\end{equation}

where the family indices i, j are summed over, the indices $i,j=1,2,3$ represent the three generations of fermions. $C=i\gamma_2\gamma_0$ is the charge conjugation 
operator, $\widetilde{\phi}=\tau_2\phi^*\tau_2$ and $\gamma_{\mu}$ are the Dirac matrices. Considering discrete parity symmetry, the Majorana Yukawa couplings $f_L=f_R$ (for left-right symmetry) gives rises
to Majorana neutrino mass after electroweak symmetry breaking  when the triplet Higgs $\Delta_L$ and $\Delta_R$ acquires non zero vev leads to $6\times6$ neutrino mass matrix which is given as-

\begin{equation}\label{eqx8}
M_\nu=\left[\begin{array}{cc}
M_{LL}&M_D\\
{M_D}^T&M_{RR}
\end{array}\right],
\end{equation}

where 
\begin{equation}\label{eqx9}
M_D=\frac{1}{\sqrt{2}}(k_1h+k_2\widetilde{h}), M_{LL}=\sqrt{2}v_Lf_L, M_{RR}=\sqrt{2}v_Rf_R,
\end{equation}

where $M_D$, $M_{LL}$ and $M_{RR}$ are the Dirac neutrino mass matrix, left-handed and right-handed mass matrix respectively. Assuming $M_L\ll M_D\ll M_R$, the 
light neutrino mass, generated within a type I+II seesaw can be written as,

\begin{equation}\label{eqx10}
M_\nu= {M_\nu}^{I}+{M_\nu}^{II},
\end{equation}

\begin{equation}\label{eqx11}
M_\nu=M_{LL}+M_D{M_{RR}}^{-1}{M_D}^T
=\sqrt{2}v_Lf_L+\frac{k^2}{\sqrt{2}v_R}h_D{f_R}^{-1}{h_D}^T,
\end{equation}

where the first and second terms in equation (\ref{eqx11}) corresponds to type-II seesaw and type-I seesaw mediated by RH neutrino respectively.
Here,
\begin{equation}\label{eqx12}
h_D=\frac{(k_1h+k_2\widetilde{h})}{\sqrt{2}k} , k=\sqrt{\left|{k_1}\right|^2+\left|{k_2}\right|^2}.
\end{equation}

In the context of LRSM both type I and type II seesaw terms can be written in terms of $M_{RR}$ which arises naturally at a high energy scale as a result of spontaneous parity breaking. In LRSM the Majorana Yukawa couplings $f_L$ and $f_R$ are the same (i.e, $f_L=f_R$) and the vev for left-handed triplet $v_L$ can be written as,

\begin{equation}\label{eqx13}
v_L=\frac{\gamma {M_W}^2}{v_R}.
\end{equation}

Thus equation (\ref{eqx11}) can be written as ,

\begin{equation}\label{eqx14}
M_\nu=\gamma(\frac{M_W}{v_R})^2M_{RR}+M_D{M_{RR}}^{-1}{M_D}^T.
\end{equation}
The dimensionless parameter $\gamma$ can be written as \cite{PhysRevD.83.035007}

\begin{equation}\label{eqx15}
\gamma=\frac{\beta_1k_1k_2+\beta_2{k_1}^2+\beta_3{k_2}^2}{(2\rho_1-\rho_3)k^2}.
\end{equation}

Here the terms $\beta$, $\rho$ are the dimensionless parameters that appear in the expression of the Higgs potential.

\section{A left right flavor symmetric model:}{\label{sec:level4}}

Symmetries play a crucial role in particle physics. Non-Abelian discrete flavor symmetries have wide applications in particle physics as these are important tools for controlling the flavor structures of the model. In the present work, the symmetry realization of the structure of the mass matrices has been carried out using the discrete flavor symmetry $A_{4}$, which is a group of permutations of four objects, isomorphic to the symmetry group of a tetrahedron. $A_{4}$ has four irreducible representations with three singlets and one triple denoted by $1$, $1^{/}$, $1^{//}$, $3_{A}$ and $ 3_{S}$ respectively. The model contains the usual particle content of LRSM. The lepton doublets transform as triplet under $A_{4}$ while  Higgs bidoublet and scalar triplets transform as $1$  under $A_{4}$. Two flavon triplet fields $\chi^{l}$ and $\chi^{\nu}$ are included in the model which transforms as triplet under  $A_{4}$. Further $Z_{2}$ symmetry is imposed to get the desired mass matrix and to constrain the non-desired interactions of the particles. Additionally, a flavon singlet $\epsilon$ is used to allow for three non-degenerate and non-zero charged lepton masses. The particle content and the charge assignments are detailed in table \ref{tab2}.

The lagrangian of all Yukawa term can be written as-
\begin{center}
	\begin{align}
 \mathcal{L_{Y}}=\bar{l_{L}}(Y_{\epsilon}\epsilon+Y_{l1}\chi^{l}+Y_{l2}\chi^{l})\Phi l_{R}+\bar{l_{L}}(\tilde{Y}_{\epsilon} \epsilon+\tilde{Y}_{l1}\chi^{l}+\tilde{Y}_{l2}\chi^{l})\tilde{\Phi}l_{R}\nonumber\\
	+\bar{l^c}_{R}(Y^0_{R} +Y^\nu_{R}\chi^{\nu})i\tau_{2}\Delta_{R} l_{R}+{\bar{l^c}_{L}}(Y^0_{R}
	+Y^\nu_{R}\chi^{\nu})i\tau_{2}\Delta_{L}l_{L}
	\end{align} \label{l1}
\end{center}

\begin{table}[h]
	\begin{center}
		\begin{tabular}{|c|c|c|c|c|c|c|c|c|c|c|}
			
			\hline 
			Field &$l_{L}$ & $l_{R}$ & $\Phi$ & $\Delta_{L}$ & $\Delta_{R}$ &  $\chi^{l}$ & $\chi^{\nu}$ & $\epsilon$ \\ 
			\hline 
			$SU(2)_{L}$ &$2$  & $1$& $2$ & $3$ & $1$& $1$ & $1$ &$1$  \\
			\hline 
			$SU(2)_{R}$ & $1$ & $2$ &$2$& $1$ & $3$ & $1$ &  $1$ &$1$ \\
			\hline
			$U(1)_{B-L}$ &$-1$&$-1$&$0$&$2$&$2$&$0$&$0$&$0$ \\
			\hline 
			$A_{4}$& $3$ & $3$ & $1$ &$1$& $1$ & $3$ & $3$ &  $1$  \\
			\hline 
			$Z_{2}$& $0$ & $0$ &$1$& $0$ & $0$ & $1$ &  $0$ &$1$\\
			\hline 
		\end{tabular} 
	\end{center}
	\caption{Fields and their respective transformations
		under the symmetry group of the model.} \label{tab2}
\end{table}

\subsection{Type-I dominance:}

In the case of type-I dominance scenario, the terms involving $\Delta_{L}$ are omitted since the vev of $\Delta_{L}$ ($<\Delta_{L}>$=$\nu_{L}$) considered to be negligible. The Yukawa matrices which are present in the eq\eqref{l1} can be written as-

\[
Y_{\epsilon} =y_{l0} \begin{pmatrix}
1  & 0 & 0 \\
0 & 1 & 0\\
0 & 0 & 1\\
\end{pmatrix}
\]
\[
Y_{l1} =y_{l1} \begin{pmatrix}
2 \chi_{1}^{l}  & -\chi_{3}^{l} & -\chi_{2}^{l} \\
-\chi_{2}^{l} & -\chi_{1}^{l} & 2 \chi_{3}^{l}\\
-\chi_{3}^{l} & 2 \chi_{2}^{l} & -\chi_{1}^{l}\\
\end{pmatrix}
\]

\[
Y_{l2} =y_{l2} \begin{pmatrix}
0  & -\chi_{3}^{l} & \chi_{2}^{l} \\
-\chi_{2}^{l} & \chi_{1}^{l} & 0\\
\chi_{3}^{l} & 0 & -\chi_{1}^{l}\\
\end{pmatrix}
\]
	Now, Dirac neutrino mass matrix $ M_{D} $ and charge lepton mass matrix $ M_l $ are given by-

\begin{equation}
M_{l}= \nu_{2} Y+ \nu_{1} \tilde{Y}, \\
M_{D}= \nu_{1} Y+ \nu_{2} \tilde{Y}
\end{equation}
Where, $\nu_{1}$ and $\nu_{2}$ are vev of the Higgs bidoublet and (Y,$\tilde{Y}$) are Yukawa coupling which is given by:

\begin{equation}
Y= Y_{\epsilon} + Y_{l1}+ Y_{l2}, \\
\tilde{Y}=\tilde{ Y_{\epsilon}} + \tilde{Y_{l1}}+\tilde{ Y_{l2}}
\end{equation}

Majorana mass matrix can be computed from-
\begin{equation}
M_{R}= \nu_{R} Y_{R} 
\end{equation}
Where $Y_{R}$ is the Majorana Yukawa coupling.

\begin{equation}
M_{R}= 
\nu_{R} y_{R0} \begin{pmatrix}
1  & 0 & 0 \\
0 & 0 & 1\\
0 & 1 & 0\\
\end{pmatrix}
+
\nu_{R} y_{R} \begin{pmatrix}
2 \chi_{1}^{\nu}  &-\chi_{3}^{\nu} &-\chi_{2}^{\nu} \\
-\chi_{3}^{\nu} & 2\chi_{2}^{\nu} &-\chi_{1}^{\nu}\\
-\chi_{2}^{\nu} & -\chi_{1}^{\nu} & 2\chi_{3}^{\nu}\\
\end{pmatrix}
\end{equation}

In our work we take the flavon alignment to be,
$ \chi^{l}-(1,0,0),\chi^{\nu}-(1,\omega,\omega^2)$. Now, this leads to diagonal Charge lepton mass matrix given by:
\[
M_{l} = \begin{pmatrix}
a+2b  & 0 & 0 \\
0 & a+(c-b) & 0\\
0 & 0 & a-(b+c)\\
\end{pmatrix}
\]

Where $a=\nu_{2}y_{l0}+\nu_{1}\tilde{y_{l0}}$,  $b=\nu_{2}y_{l1}+\nu_{1}\tilde{y_{l1}}$
and $c=\nu_{2}y_{l2}+\nu_{1}\tilde{y_{l2}}$.

Now, the Dirac neutrino mass matrix($M_{D}$) can be simplified into the form given below-
\[
M_{D} =\lambda \begin{pmatrix}
1  & 0 & 0 \\
0 & r_{1} & 0\\
0 & 0 & r_{2}\\
\end{pmatrix}
\]
where $\lambda=\nu_{1}y_{l0}+\nu_{2}\tilde{y_{l0}}$,  $r_{1}=\frac{\nu_{1}y_{l1}+\nu_{2}\tilde{y_{l1}}}{\lambda}$ and $r_{2}=\frac{\nu_{1}y_{l2}+\nu_{2}\tilde{y_{l2}}}{\lambda}$

The Majorana mass mtrix is:
\[
M_{R} =a_{R} \begin{pmatrix}
2z+1  & -\omega^2 z & -\omega z\\
-\omega^2 z & 2\omega z & 1-z\\
-\omega z & 1-z & 2 \omega^2 z\\
\end{pmatrix}
\]
Where $a_{R}=\nu_{R}y_{R0}$ and $z=\frac{y_{R}}{y_{R0}}$. Now, the relevant mass generation formula for type-I dominance:

\begin{equation}
m_{\nu}=M_{D}^T M_{R}^{-1} M_{D}
\end{equation}

So, the light neutrino mass matrix will be:
\[
m_{\nu} =\frac{m}{3z+1} \begin{pmatrix}
z+1  & \omega z r_{1} & \omega^2 z r_{2}\\
\omega z r_{1} & \omega^2 \frac{z(3z+2)r_{2}^2}{3z-1} & \frac{(z-3z^2+1)r_{1}r_{2}}{1-3z}\\
\omega^2 z r_{2} &\frac{(z-3z^2+1)r_{1}r_{2}}{1-3z} & \frac{\omega z(3z+2)r_{2}^2}{3z-1}\\
\end{pmatrix}
\]

\subsection{Type-II dominance: }

In Type-II we have to consider the terms involving $\Delta_{L}$ of the Lagrangian. To break the $\mu-\tau$ symmetry of resulting mass matrix we need to introduce another flavon $\epsilon ^{'}$. This flavon transforms as $1^{'}$ under $A_{4}$. Now lagrangian for the neutrino sector will become-
\begin{center}
	\begin{align}
	\mathcal{L_{Y \nu}}=
	\bar{l^c}_{R}(Y^0_{R} \epsilon^{'} +Y^\nu_{R}\chi^{\nu})i\tau_{2}\Delta_{R} l_{R}+{\bar{l^c}_{L}}(Y^0_{R} \epsilon^{'}
	+Y^\nu_{R}\chi^{\nu})i\tau_{2}\Delta_{L}l_{L}
	\end{align}
\end{center}

So, we can write-

\begin{equation}
m_{\nu}=\nu_{L}Y_{L}  ,  \\ 
M_{R}= \nu_{R} Y_{R}
\end{equation}

Where $\nu_{L}$ and $\nu_{R}$ are VEV of $\Delta_{L}$ and $\Delta_{R}$ respectively. $Y_{L}$ and $Y_{R}$ are Majorana Yukawa couplings which are taken to be equal in the scheme of LRSM.
Now we know that the Majorana mass matrix is given by:

\begin{equation}
M_{R}= 
\nu_{R} y_{R0} \begin{pmatrix}
0  & 0 & 1 \\
0 & 1 & 0\\
1 & 0 & 0\\
\end{pmatrix}
+
\nu_{R} y_{R} \begin{pmatrix}
2 \chi_{1}^{\nu}  &-\chi_{3}^{\nu} &-\chi_{2}^{\nu} \\
-\chi_{3}^{\nu} & 2\chi_{2}^{\nu} &-\chi_{1}^{\nu}\\
-\chi_{2}^{\nu} & -\chi_{1}^{\nu} & 2\chi_{3}^{\nu}\\
\end{pmatrix}
\end{equation}

Using the chosen flavon alignment we get-

\[
M_{R} =a_{R} \begin{pmatrix}
2z  & -\omega^2 z & 1 -\omega z\\
-\omega^2 z &1+ 2\omega z & -z\\
1-\omega z & -z & 2 \omega^2 z\\
\end{pmatrix}
\]
Where $a_{R}=\nu_{R}y_{R0}$ and $z=\frac{y_{R}}{y_{R0}}$

Similarly, we can compute the light neutrino mass matrix which is given by-

\[
m_{\nu} =a_{L} \begin{pmatrix}
2z  & -\omega^2 z & 1 -\omega z\\
-\omega^2 z &1+ 2\omega z & -z\\
1-\omega z & -z & 2 \omega^2 z\\
\end{pmatrix}
\]

Where $a_{L}=\nu_{L}y_{R0}$ and $z=\frac{y_{R}}{y_{R0}}$
\section{NEutrinoless double beta decay(NDBD)in LRSM:}{\label{sec:level5}}

 In the scheme of LRSM \cite{Mohapatra:1979ia, Senjanovic:1975rk, Mohapatra:1979ia, Pati:1974yy}, due to the presence of the heavy scalar particles, NDBD receives additional contributions. Violation of lepton number can be manifested in neutrinoless double beta decay process. The phenomenological importance of NDBD  in neutrino physics is vey high. We will study NDBD within the framework of LRSM which is realised by using $A_{4}\times Z_{2}$. Many of the erlier work \cite{Picciotto:1982qe,Ge:2015yqa,Hirsch:1996qw,Tello:2010am,BhupalDev:2013ntw,Chakrabortty:2012mh} has been done on NDBD within the framework of LRSM. NDBD mediated by the light Majorana neutrinos, the effective mass governing the process is given by,
\begin{equation}
	 m_{\nu}^{eff}= U_{Li}^2 m_{i} \label{eq2}
\end{equation}
where, $ U_{Li}$ are the elements of the first row of the neutrino mixing matrix, $U_{PMNS}$,which is dependent on known parameters  $\theta_{13}$,$\theta_{12}$  and the unknown Majorana phases $\alpha$ and $\beta$.  $ U_{PMNS}$ is the diagonalizing matrix of the light neutrino mass matrix, $m_{\nu}$
such that,
\begin{equation}
	 m_{\nu}=U_{PMNS}  M_{\nu}^{(diag)}  U_{PMNS}^T
\end{equation}
where, $ M_{\nu}^{(diag)} =diag(m_{1},m_{2},m_{3})$ and,
\begin{equation}
	 U=\begin{bmatrix} c_{12} c_{13}& s_{12} c_{13}& s_{13} e^{-i \delta}  \\ -c_{23} s_{12}-s_{23} s_{13} c_{12} e^{i\delta}  & -c_{23} c_{12}-s_{23} s_{12} s_{13} e^{i\delta}& s_{23} c_{13}\\ s_{23} s_{12}-c_{23} s_{13} c_{12} e^{i\delta}& -s_{23} c_{12}-c_{23} s_{13} s_{12} e^{i\delta}& c_{23} c_{13}\end{bmatrix} P
\end{equation}
where $P$ contains majorana phases $\alpha$ and $\beta$ and $P=diag(1,e^{i\alpha},e^{i\beta})$.
We can parameterize the effective Majorana mass in terms of the elements of diagonalizing matrix and the mass eigen values as, 
\begin{equation}
	 m_{\nu}^{eff}=m_{1} c_{12}^2 c_{13}^2+m_{2} s_{12}^2 c_{13}^2 e^{2i\alpha} + m_{3}s_{13}^2 e^{2i\beta} 
\end{equation}

In the context of LRSM, there are several contributions to NDBD in addition to the standard contribution via light Majorana neutrino exchange owing to the presence of several heavy additional scalars, vector and, fermionic fields. Many of the earlier works have explained it in details \cite{Dev:2014xea,Bambhaniya:2015ipg,Awasthi:2016kbk,Borgohain:2017akh}.  The various contributions to NDBD decay transition rate in LRSM are briefly summarized below.

1) Standard Model contribution to NDBD where the intermediate particles are the $W_{L}^-$ bosons and light neutrinos. The amplitude of this process depends upon the leptonic mixing matrix elements and light neutrino masses.

2) Heavy right-handed neutrino contribution to NDBD in which the mediator particles are the $W_{L}^-$ bosons. The amplitude of this process depends upon the mixing between light and heavy neutrinos as well the mass of the heavy neutrinos.

\vspace{1cm}

%\hspace{0.5cm} \includegraphics[width=0.7\textwidth]{WL.eps}

%{Fig:1-Neutrinoless double beta decay contribution from light and heavy Majorana neutrinos
%	from two $W_{L}$ exchange}

%\vspace{0.5cm}

%Amplitude of the contribution from the above two diagram are respectively-
%\begin{equation}
%\mathcal{A}_{\nu}^{LL}\propto \frac{1}{M_{W_{L}}^4} \frac{U_{L_{ei}}^2 m_{i}}{p^2}
%\end{equation}

%\begin{equation}
%\mathcal{A}_{N}^{LL}\propto \frac{1}{M_{W_{L}}^4} \frac{T_{ei}^2 M_{i}}{p^2}
%\end{equation}
3) Light neutrino contribution to NDBD in which the intermediate particles are $W_{R}^-$ bosons. The amplitude of this process depends upon the mixing between light and heavy neutrinos as well as the mass of the right handed gauge boson, $W_{R}^-$ boson.

4) Heavy right-handed neutrino contribution to NDBD in which the mediator particles are the $W_{R}^-$ bosons. The amplitude of this process depends upon the elements of the right-handed leptonic mixing matrix and the mass of the right-handed gauge boson, $W_{R}^-$ boson as well as the mass of the heavy right-handed Majorana neutrinos.

%\vspace{1cm}

%\hspace{0.5cm} \includegraphics[width=0.7\textwidth]{WR}

%{Fig:2-Neutrinoless double beta decay contribution from light and heavy Majorana neutrinos
%	from two $W_{R}$ exchange}
%\vspace{0.5cm}

%Amplitude of the contribution from the above two diagram are respectively-

%\begin{equation}
%\mathcal{A}_{\nu}^{RR}\propto \frac{1}{M_{W_{R}}^4} \frac{S_{ei}^2 m_{i}}{p^2}
%\end{equation}

%\begin{equation}
%\mathcal{A}_{N}^{RR}\propto \frac{1}{M_{W_{R}}^4} \frac{U_{R_{ei}}^2}{M_{i}}
%\end{equation}

5) Light neutrino contribution from the Feynman diagram mediated by both $W_{L}^-$ and $W_{R}^-$. The amplitude of this process depends upon the mixing between light and heavy neutrinos, leptonic mixing matrix elements, light neutrino masses, and the mass of the gauge bosons, $W_{L}^-$ and $W_{R}^-$.

6) Heavy neutrino contribution from the Feynman diagram mediated by both $W_{L}^-$ and $W_{R}^-$. The amplitude of the process depends upon the right-handed leptonic mixing matrix elements, mixing between the light and heavy neutrinos as well as the mass of the gauge bosons,$W_{L}^-$ and $W_{R}^-$ and the mass of the heavy right-handed neutrinos.

\vspace{1cm}

%\hspace{0.5cm} \includegraphics[width=0.7\textwidth]{WL-WR}

%{Fig:3-Neutroless double beta decay contribution from light and heavy Majorana neutrino intermediate states from both left and right handed gauge bosons exchange at each vertices.}
%\vspace{0.5cm}

%Amplitude of the contribution from the above two diagram are respectively-

%\begin{equation}
%\mathcal{A}_{\nu}^{LR}\propto \frac{1}{M_{W_{R}}^4 M_{W_{L}}^4 } \frac{S_{ei} U_{L_{ei}}m_{i}}{p^2}
%\end{equation}

%\begin{equation}
%\mathcal{A}_{N}^{LR}\propto \frac{1}{M_{W_{R}}^4 M_{W_{L}}^4} \frac{U_{R_{ei}} T_{ei}}{M_{i}}
%\end{equation}
7) Triplet Higgs $\bigtriangleup _{L}$ contribution to NDBD in which the mediator particles are $W_{L}^-$ bosons. The amplitudes for the process depends upon the masses of the $W_{L}^-$ bosons, left handed triplet Higgs, $\bigtriangleup _{L}$ as well as their coupling to leptons, $f_{L}$.

8) Right-handed triplet Higgs $\bigtriangleup_{R}$ contribution to NDBD in which the mediator particles are $W_{R}^-$ bosons. The amplitude for the process depends upon the masses of the $W_{R}^-$ bosons, right-handed triplet Higgs, $\bigtriangleup_ {R}$ as well as their coupling to leptons, $f_{R}$.

\vspace{1cm}

%\hspace{0.5cm} \includegraphics[width=0.7\textwidth]{DELTA}

%{Fig4:Neutrinoless double beta decay contribution from the charged Higgs intermediate states
%	from $W_{L}$ and $W_{R}$ exchange.}
%\vspace{0.5cm}

%Amplitude of the contribution from the above two diagram are respectively-

%\begin{equation}
%\mathcal{A}_{\bigtriangleup_{L}}^{LR}\propto \frac{1}{M_{W_{L}}^4 M_{\bigtriangleup_{L}}^4 } f_{L} v_{L}
%\end{equation}

%\begin{equation}
%\mathcal{A}_{\bigtriangleup_{R}}^{RR}\propto \frac{1}{M_{W_{R}}^4 M_{\bigtriangleup_{R}}^4 } f_{R} v_{R}
%\end{equation}

However, in this work, we have considered only three of the above-mentioned
contributions to NDBD. One from the standard light neutrino contribution through the exchange
of $W^{-}_{L}$  and the other two are the new physics contributions to NDBD that are the ones mediated by $W^{-}_{R}$ and $\Delta_{R}$ respectively.
For simple
approximations, an assumption of similar mass scales for the heavy particles has been made in the LRSM, where, $M_{R} \approx  M_{W_{R}} \approx M_{\Delta^{++}_{L}} \approx
M_{\Delta_{R}^{++}} \approx TeV$, at a scale accessible at
the LHC. Under these assumptions, the amplitude for the light-heavy mixing contribution
which is proportional to $\frac{m_{D}^2}{M_{R}}$ remains very small (since $m_{\nu} \approx \frac{m_{D}^2}{M_{R}} \approx (0.01 - 0.1)eV$,
$m_{D} \approx (10^5 - 10^6) eV$ which implies $\frac{m_{D}}{M_{R}} \approx (10^{-7} - 10^{-6})$ eV. Thus, we ignore the
contributions involving the light and heavy neutrino mixings. For a simplified approach, we have also ignored the mixing between $W_{L}$ and $W_{R}$ bosons owing to the above mentioned assumptions, which would cause a further supression in the amplitude of the process. However,the $\Delta_{R}$ mediated diagram can in principle contribute for $W_{R}$ mass around TeV scale.
But invoking the constrain from Lepton Flavour Violating(LFV) decays it is seen that for the majority of the parameter space $\frac{M_{N}}{M_{\Delta}} < 0.1$ and hence the $\Delta_{R}$ contribution can be
suppressed. We will consider for the case where $M_{N} \approx M_{\Delta}$.
 In the next section, we have discussed lepton flavor violation in the framework of LRSM and after that, we present a detailed analysis of
our work and we have divided it into different subsections, firstly the standard light neutrino contribution to NDBD and then the new physics contribution to NDBD considering both type-I and then type-II dominant cases. And we have also tested our model by incorporating LFV constraints coming from different relevant experiments.

\section{Lepton flavor violation:}
Theoretical and experimental manifestation of LFV \cite{Dolan:2018yqy, BhupalDev:2014qbx, Nemevsek:2012iq} has been one of the most promising areas of research for a long. 
$\mu\rightarrow e\gamma$, $\mu\rightarrow 3e$, and $\mu\rightarrow e$ conversion in the nuclei are the most prominent low energy LFV channels which are accessible in recent experiments. The relevant branching ratio(BR) \cite{Borgohain:2017akh} of the above mentioned processes are given below-
\begin{equation}
BR_{\mu\rightarrow e\gamma}=\frac{\Gamma(\mu^{+}\rightarrow e^{+}\gamma)}{\Gamma_{\mu}}
\end{equation} 
\begin{equation}
BR_{\mu\rightarrow e}^{Z}=\frac{\Gamma(\mu^{-}+A(N,Z)\rightarrow e^{-}+A(N,Z)}{\Gamma_{capt}^{Z}}
\end{equation} 
\begin{equation}
BR_{\mu\rightarrow 3e}=\frac{\Gamma(\mu^{+}\rightarrow e^{+}e^{-}e^{+})}{\Gamma_{\nu}}
\end{equation}
 In this model, we considered these processes. The constraint on branching
ratio of the process $\mu\rightarrow 3e$  is $ <1.0 \times 10^{-12}$ given by the SINDRUM experiment \cite{Perrevoort:2017uex}. Mu3e collaboration \cite{Mu2e} has intended to improve this limit by four orders. The MEG collaboration \cite{TheMEG} has put the upper bound on the branching ratio of the decay $\mu\rightarrow e\gamma$ to be $ <4.2 \times  10^{-13}$.Taking into account the contributions from heavy right-handed neutrinos and Higgs scalars, the expected branching ratios and conversion rates of the above processes have been calculated in the LRSM in this work.

Branching ratio for the process $\mu\rightarrow 3e$ is given by-
\begin{equation}
BR_{\mu\rightarrow 3e}=\frac{1}{2}|h_{\mu e} h_{ee}^{*}|^{2} \bigg(\frac{m_{W_{L}^{4}}}{M^{4}_{\Delta_{L}^{++}}}+\frac{m_{W_{R}^{4}}}{M^{4}_{\Delta_{R}^{++}}}\bigg) \label{lfv1}
\end{equation} 
where $h_{ij}$ is the lepton Higgs coupling in the LRSM, which is given by-
\begin{equation}
h_{ij}= \sum_{n=1}^{3} V_{in} V_{jn} \bigg(\frac{M_{n}}{M_{W_{R}}}\bigg), i,j=e,\mu,\tau \label{lfv2}
\end{equation}
Now, in the case of $\mu\rightarrow e\gamma$ process, branching ratio is given by-
\begin{equation}
BR_{\mu\rightarrow e\gamma}=1.5 \times 10^{-7}|g_{lfv}|^{2} \bigg(\frac{1TeV}{M_{W_{R}}}\bigg)^{4} \label{lfv3}
\end{equation}
where $g_{lfv}$ is defined as-
\begin{equation}
g_{lfv}=\sum_{n=1}^{3} V_{\mu n} V_{en}^{*} \big(\frac{M_{n}}{M_{W_{R}}}\big)^{2}=\frac{\bigg[M_{R} M_{R}^{*}\bigg]_{\mu e}}{M_{W_{R}}}
\end{equation}
The sum is over heavy neutrino. V is the right-handed neutrino mixing matrix and $M_{\Delta_{L,R}}^{++}$ are the mass of doubly charged boson.

\section{Numerical analysis and Results: }

In our present work, we have constructed a flavor symmetric model for both type-I and type-II dominance and studied LNV (NDBD) for standard as well as non-standard
contributions for the effective mass as well as the half-life governing the decay process along with different LFV processes in the framework of LRSM. We also checked the consistency of the model by varying different neutrino oscillation parameters with the light neutrino contribution to the effective mass coming from the model for both type-I and type-II dominant cases. In this section, we present a detailed analysis of
our work and we have divided it into different subsections, firstly the standard light neutrino contribution to NDBD and then the new physics contribution to NDBD considering both type II and then type I dominance case. We have also studied lepton flavor violating processes such as $\mu \rightarrow 3e$ and $\mu \rightarrow e \gamma$ and correlated with neutrino mass within the model.

\subsection{Standard light neutrino contribution:}

For NDBD mediated by the light Majorana neutrinos the effective mass governing the process is as given in (\ref{eq2}). In our present
work, we first evaluated the effective light neutrino mass within the standard mechanism
using the formula (\ref{eq2}) where $U_{Li}$ are the elements of the first row of the neutrino mixing
matrix.
$U_{PMNS}$ is the diagonalizing matrix of the light neutrino mass matrix, $m_{\nu}$,
such that-
\begin{equation}\label{eqa4}
m_{\nu}= U_{PMNS}{M_\nu}^{(diag)} {U_{PMNS}}^T
\end{equation}
Where $M_{\nu}^{diag}=diag(m_{1},m_{2},m_{3})$. In case of three neutrino mixing we can have two neutrino mass spectra.

1) Normal Hierachy (NH) which corresponds to $m_{1}<m_{2}<<m_{3}$;$\Delta m_{12}^2<<\Delta m_{23}^2$

1) Inverted Hierachy (IH) which corresponds to $m_{3}<<m_{1}\approx m_{3}$;$\Delta m_{12}^2<<|\Delta m_{13}^2|$

In both spectra, $\Delta m_{12}^2=\Delta m_{solar}^2$. For NH, $\Delta m_{23}^2=\Delta m_{atm}^2$ and for IH,$|\Delta m_{13}^2|=\Delta m_{atm}^2$

In the case of NH, the neutrino masses $m_{2}$ and $m_{3}$ are connected with
the lightest mass $m_{1}$ by the relation,

\begin{equation}
m_{2}=\sqrt{m_{1}^2+\Delta m_{solar}^2}     ;  \nonumber 
m_{3}=\sqrt{m_{1}^2+\Delta m_{solar}^2+\Delta m_{atm}^2}
\end{equation}

In the case of IH,
the lightest mass is $m_{3}$ and we have,

\begin{equation}
m_{1}=\sqrt{m_{3}^2+\Delta m_{atm}^2}     ;  \nonumber 
m_{2}=\sqrt{m_{3}^2+\Delta m_{solar}^2+\Delta m_{atm}^2}
\end{equation}

\begin{figure*}
	\begin{center}
		\includegraphics[width=0.42\textwidth]{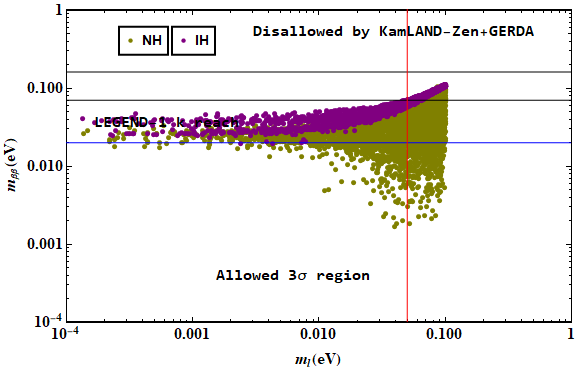}
		\includegraphics[width=0.40\textwidth]{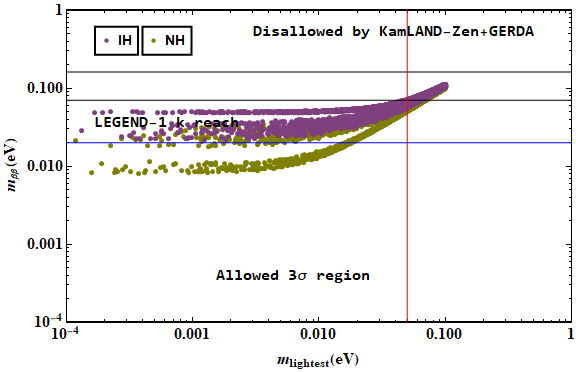}

	\end{center}	
	\begin{center}
		\caption{:The light neutrino contribution to  neutrinoless double beta decay process for typeI(left) and typeII(right) considering both NH and IH cases. The band of two
			black solid line and the red solid line represents the KamLAND-Zen bound on the effective mass and
			the Planck bound on the sum of the absolute neutrino mass respectively. And, the blue line reperents the future sensitivity on effective mass in Legend-1k reach \cite{LEGEND:2017cdu}.}
		\label{fig10}
	\end{center}
\end{figure*}

We have computed the light neutrino mass matrix from the model described at the beginning for both type-I and type-II cases, which are-

\[
 m_{\nu}(type-I) =\frac{m}{3z+1} \begin{pmatrix}
z+1  & \omega z r_{1} & \omega^2 z r_{2}\\
\omega z r_{1} & \omega^2 \frac{z(3z+2)r_{2}^2}{3z-1} & \frac{(z-3z^2+1)r_{1}r_{2}}{1-3z}\\
\omega^2 z r_{2} &\frac{(z-3z^2+1)r_{1}r_{2}}{1-3z} & \frac{\omega z(3z+2)r_{2}^2}{3z-1}\\
\end{pmatrix}
\]

\[
m_{\nu}(type-II) =a_{L} \begin{pmatrix}
2z  & -\omega^2 z & 1 -\omega z\\
-\omega^2 z &1+ 2\omega z & -z\\
1-\omega z & -z & 2 \omega^2 z\\
\end{pmatrix}
\]

\begin{figure*}
	\begin{center}
		\includegraphics[width=0.40\textwidth]{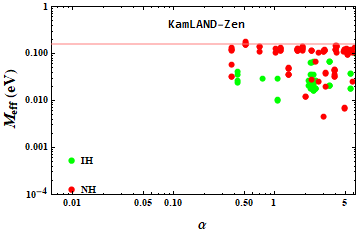}
		\includegraphics[width=0.40\textwidth]{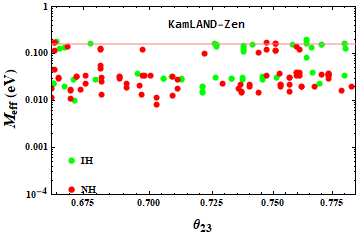}
		\includegraphics[width=0.40\textwidth]{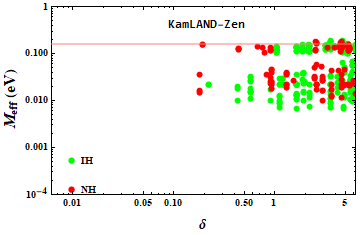}
		
	\end{center}	
	\begin{center}
		\caption{ Variation of $\alpha$, $\theta_{23}$ and $\delta$ with effective mass for type-II dominance case.}
		\label{fig70}
	\end{center}
\end{figure*}

\begin{figure*}
	\begin{center}
		\includegraphics[width=0.40\textwidth]{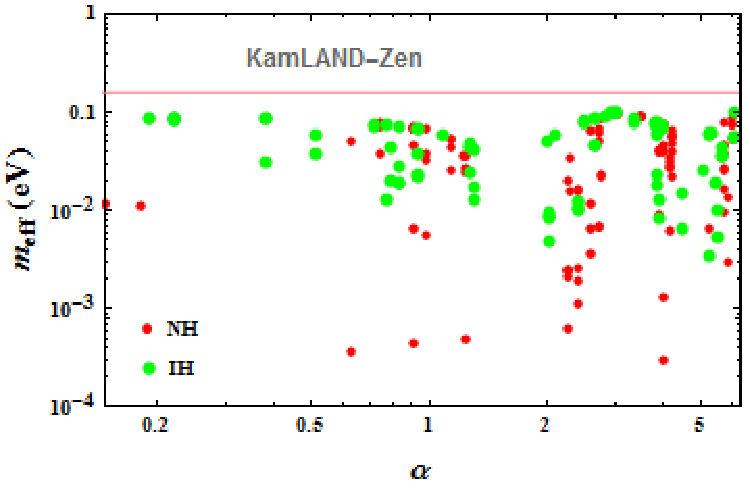}
		\includegraphics[width=0.40\textwidth]{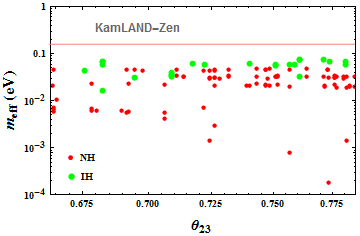}
		\includegraphics[width=0.40\textwidth]{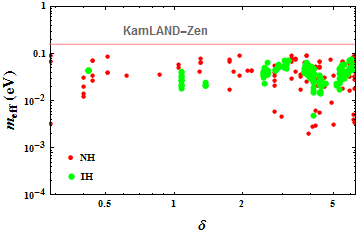}
		
	\end{center}	
	\begin{center}
		\caption{ Variation of $\alpha$,$\theta_{23}$ and $\delta$ with effective mass for type-I dominance case.}
		\label{fig1}
	\end{center}
\end{figure*}

As discussed before, the structures of the different matrices involved are formed using the discrete flavor symmetry $A_{4}\times Z_{2}$ and we obtain the resulting light neutrino mass matrix. The light neutrino mass matrix arising from the model is consistent with non-zero $\theta_{13}$ as $A_{4}$ product rules lead to the light neutrino mass matrix in which the $\mu-\tau$ symmetry is explicitly broken.
 Using the $3\sigma$ \cite{Esteban:2020cvm} ranges of the neutrino oscillation parameters we solve for the different model parameters of the model. Then we calculated the effective mass for both cases. The effective mass assumes
different values depending on whether the neutrino mass states follow a normal hierarchy
(NH) or inverted hierarchy (IH). The variation is shown in Fig\ref{fig10}. It is seen from the
figure that the light neutrino contribution to the NDBD process can
saturate the bound imposed by KamLAND-ZEN.

In light of standard light neutrino contribution to the effective mass, we varied different neutrino oscillation parameters to check the viability of the model. In Fig\ref{fig1} and Fig\ref{fig70}, we have shown different variational plots of neutrino oscillation parameters with effective mass for type-I and type-II dominant cases. From these plots, we can say that the parameters Majorana phase $\alpha$, mixing angle $\theta_{23}$ and CP-violating phase $\delta$ are well within the experimental limits.

%\begin{figure}[h!]
	%\includegraphics[width=0.4\textwidth,height=3.88cm]{light1}
	%\includegraphics[width=0.4\textwidth,height=3.88cm]{light2}
%\end{figure}

\subsection{New physics contribution to NDBD:}

We have also considered the contribution to NDBD from the right-handed current and triplet Higgs($\Delta_{R}$). Although the contribution of the $\Delta_{R}$ can be suppressed if we invoke the constraints from LFV decays. We will discuss this contribution in certain conditions. 

The contribution coming from right-handed current can be written as-

\begin{equation}\label{eq55}
 {m_{N}^{\beta\beta}=p^2\frac{{M_{W_L}}^4}{{M_{W_R}}^4}\frac{{U_{Rei}}^*2}{M_i}}
\end{equation}

\par Here, $<p^2> = m_e m_p \frac{\mathcal{M_N}}{\mathcal{M_{\nu}}}$ is the typical momentum exchange of the process, where $ m_p$ and $ m_e$ are the mass of the proton and electron respectively and $\mathcal{M_N}$ is the NME corresponding to the RH neutrino exchange. We know that TeV scale LRSM plays an important role in 0$\nu\beta\beta$ decay. We have considered the values $ M_{W_R}$ = 10 TeV, $ M_{W_L}$ = 80 GeV, $ M_{\Delta_R}\approx $3TeV, the heavy RH neutrino $\approx$ TeV which are within the recent collider limits. 
The allowed value of $p$, the virtuality of the exchanged neutrino is in the range $\sim $ (100-200) MeV and we have considered $p\simeq$180 MeV.

\par Thus,
\begin{equation}\label{eq56}
 p^2\frac{{M_{W_L}}^4}{{M_{W_R}}^4} \simeq 10^{10} {eV}^2.
\end{equation}
However, equation (\ref{eq56}) is valid only in the limit ${M_i}^2 \gg\left|<p^2>\right|$ and $ {M_\Delta}^2\gg\left|<p^2>\right|$. 

Under the above approximations the time-period for $0\nu\beta\beta$ process can be written as:, 
\begin{equation}
\Gamma^{0\nu}=G^{0\nu}(Q,Z)(|\mathcal{M}_{\nu}|^2 |\eta_{L}|^2+|\mathcal{M}_{N}|^2 |\eta_{R}|^2)
\end{equation}
Where,
\begin{equation}
 |\eta_{L}|=\frac{|U_{Lei}^2 m_{i}|}{m_{e}}=\frac{m_{\nu}^{\beta\beta}}{m_{e}}
\end{equation}

\begin{equation}
	 |\eta_{R}|=\frac{{M_{W_L}}^4}{{M_{W_R}}^4} |\frac{U_{Rei}^{*2}}{M_i}|
\end{equation}
The contribution from the neutrino propagator term to the amplitude
is $\approx \frac{m_{i}}{p^2-m_{i}^2}$
The different dependence on the masses for the left and right sector come since
the exchanged momentum p satisfies $m_{i} << p << M_{i}$
. $\mathcal{M_{\nu}}$ and  $\mathcal{M_{N}}$ are the nuclear matrix
elements corresponding to light and heavy neutrino exchange respectively.
So, the time period governing the NDBD process can be given by
\begin{equation}
 \Gamma^{0\nu}=\frac{1}{{T_{\frac{1}{2}}}^{0\nu}}=G^{0\nu}(Q,Z){\left|M^{0\nu}\right|}^2\frac{{\left|m_{N+\nu}^{eff}\right|}^2}{{m_e}^2}.
\end{equation}
where
\begin{equation}\label{eq36}
 {\left| {m_{N+\nu}}^{eff}\right|}^2={\left|{m_N}^{\beta\beta}+{m_{\nu}}^{\beta\beta}\right|}^2.
\end{equation}

\begin{figure*}[h]
	\begin{center}
		\includegraphics[width=0.42\textwidth]{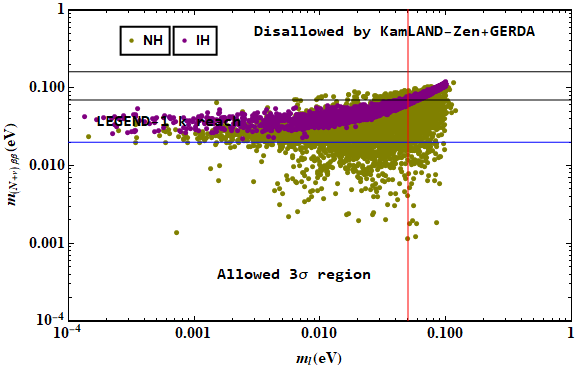}
		\includegraphics[width=0.40\textwidth]{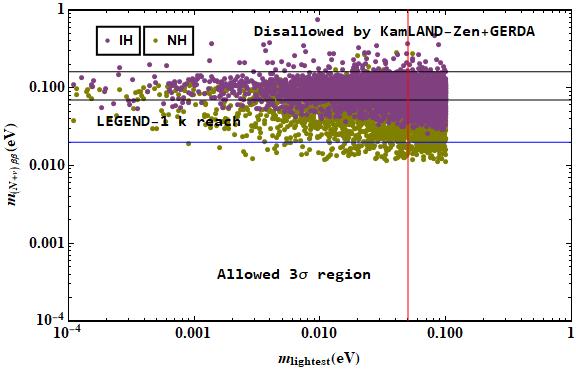}

	\end{center}	
	\begin{center}
		\caption{: The total contribution  to  neutrinoless double beta decay process considering new physics contribution coming from heavy neutrino i.e $|{m_{N+\nu}}^{eff}|$ for type-I(left) and type-II(right) considering both NH and IH cases. The band of two
				black solid line and the red solid line represents the KamLAND-Zen bound on the effective mass and
				the Planck bound on the sum of the absolute neutrino mass respectively. And, the blue line reperents the future sensitivity on effective mass in Legend-1k reach.}
		\label{fig2}
	\end{center}
\end{figure*}

\begin{figure*}
	\begin{center}
		\includegraphics[width=0.40\textwidth]{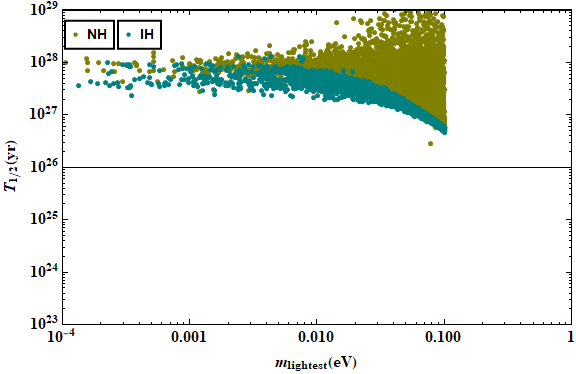}
		\includegraphics[width=0.40\textwidth]{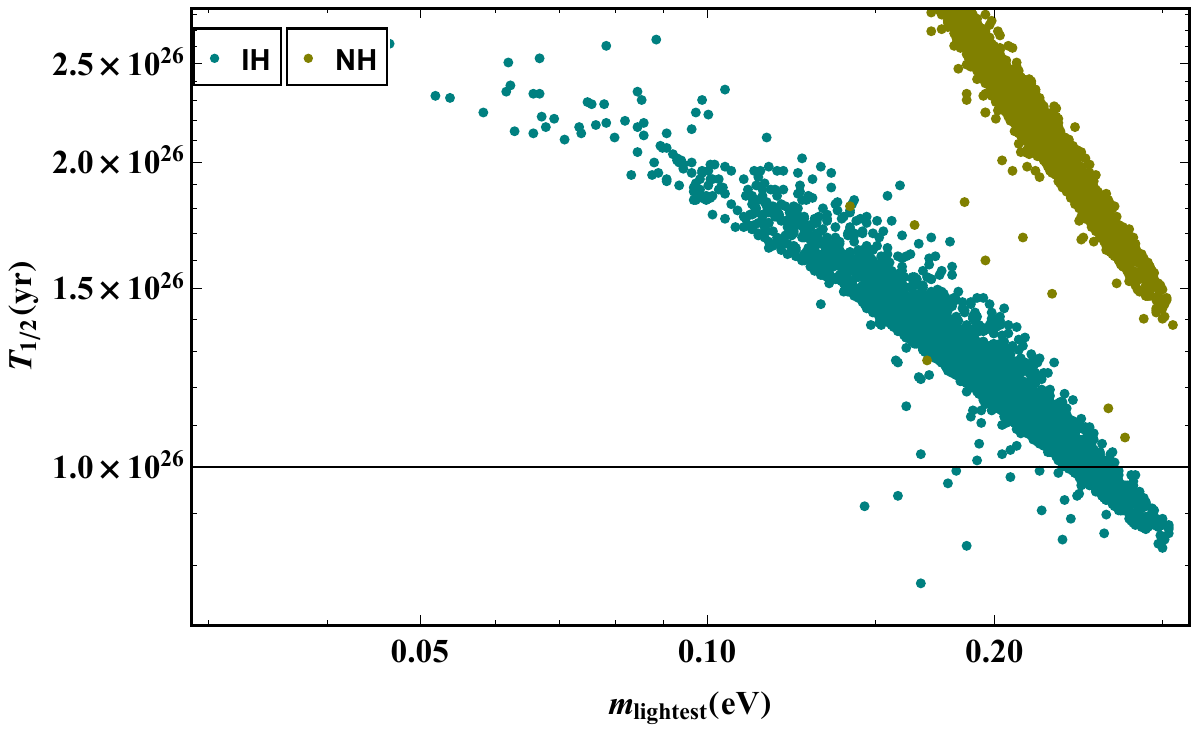}

	\end{center}	
	\begin{center}
		\caption{: The new physics contribution to half-life of  neutrinoless double beta decay process for typeI(left) and typeII(right) considering both NH and IH case. The horizontal line represents the KamLAND-Zen lower bound on the half-life of NDBD.}
		\label{fig30}
	\end{center}
\end{figure*}

To evaluate ${m_{(N+\nu)}}^{eff}$, we need the diagonalizing matrix of the heavy right-handed Majorana mass matrix $ M_{R}$, $ U_{Rei}$ and its mass eigenvalues, $ M_{i}$. We have computed the right-handed neutrino mass matrix from the model described for both type-I and type-II cases. Using the values of the model parameters, we evaluated the right-handed current contribution to the NDBD. From this, we calculated the total effective mass for the NDBD process. Variation of lightest neutrino mass with the total new contribution to effective mass and half-life of NDBD process are given in Fig\ref{fig2} and Fig\ref{fig30} for type-I and type-II dominant cases respectively.

\begin{figure*}[h]
	\begin{center}
		\includegraphics[width=0.42\textwidth]{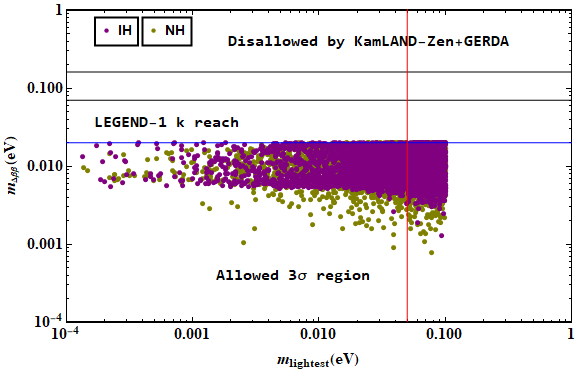}
		\includegraphics[width=0.40\textwidth]{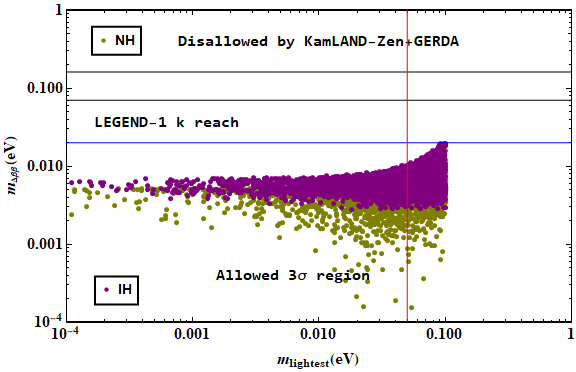}

	\end{center}	
	\begin{center}
		\caption{:The scalar triplet contribution to  neutrinoless double beta decay process for typeI(left) and typeII(right) considering both NH and IH case.The band of two
				black solid line and the red solid line represents the KamLAND-Zen bound on the effective mass and
				the Planck bound on the sum of the absolute neutrino mass respectively. And, the blue line reperents the future sensitivity on effective mass in Legend-1k reach.} 
		\label{fig4}
	\end{center}
\end{figure*}

\subsection{Scalar triplet contribution to NDBD:}

The Majorana masses of light and heavy neutrinos come naturally in the left-right model because of the two triplets $\Delta_{L, R}$. The contribution from $\Delta_{L}$ is much suppressed as compared to the dominant contributions. However the magnitude of the $\Delta_{R}$ contribution is controlled by the factor $\frac{M_{i}}{M_{\Delta_{R}}}$. In the total contribution, we
have not included the contribution due to the triplet Higgs contribution under the assumption $\frac{M_{i}}{M_{\Delta_{R}}}<0.1$, which is obtained from LFV processes. However, this approximation
though valid in a large part of the parameter space there are some allowed mixing parameters for which this ratio can be higher. In that case, we need to include this contribution. We discuss the impact of this contribution in the limit, $ M_{\Delta_{R}}=M_{heaviest}$. Now, we can write down the contribution of scalar triplet($\Delta_{R}$) to the effective mass as-
\begin{equation}
 |m_{\Delta^{ee}}|= |p^2\frac{{M_{W_L}}^4}{{M_{W_R}}^4} \frac{2 M_{N}}{M_{\Delta_{R}}}|
\end{equation}
We have evaluated the contribution from scalar triplet to the NDBD process and plotted the contribution of the effective mass due to the triplets with the lightest neutrino mass for
type-I and type-II seesaw cases which are given in the Fig\ref{fig4}.

\begin{figure}[h]
	\begin{center}
		\includegraphics[width=0.42\textwidth]{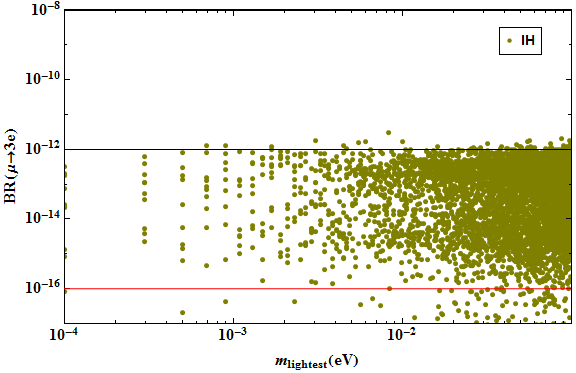}
		\includegraphics[width=0.40\textwidth]{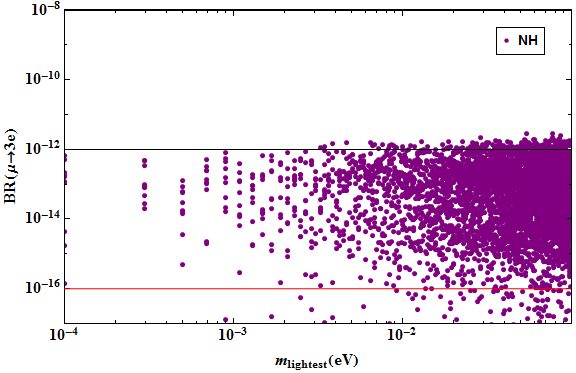}\\
		\includegraphics[width=0.42\textwidth]{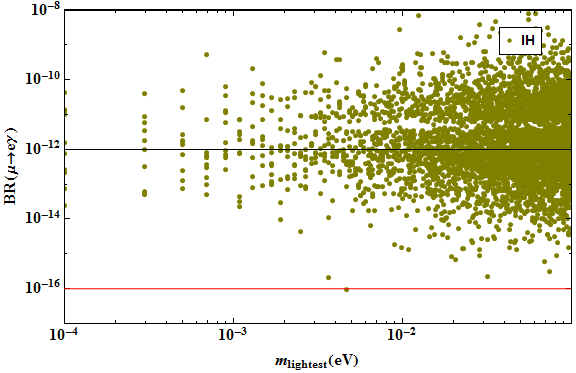}
		\includegraphics[width=0.40\textwidth]{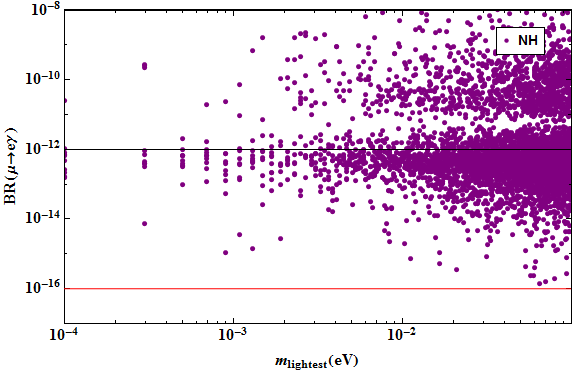}
		
	\end{center}	
	\begin{center}
		\caption{Total contribution to lepton flavour violation shown as a function of the lightest neutrino mass for in case of type-I dominance case for both  $\mu\rightarrow 3e$ and  $\mu\rightarrow e\gamma$  .The blue and red horizontal line
			shows the limit of BR as given by SINDRUM experiment and the recently proposed limit of $\mu\rightarrow 3e$ experiment
			respectively} 
		\label{fig79}
	\end{center}
\end{figure}

\subsection{Correlating LFV and neutrino mass}

\begin{figure}[h]
	\begin{center}
		\includegraphics[width=0.42\textwidth]{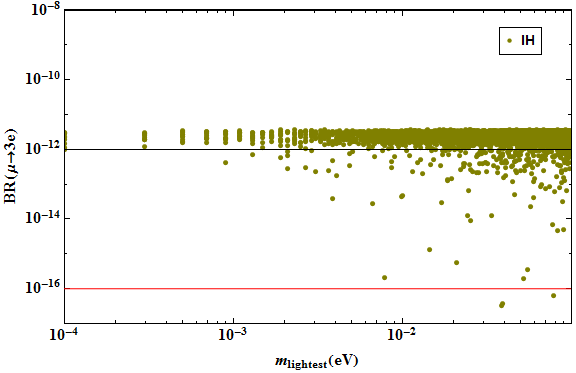}
		\includegraphics[width=0.40\textwidth]{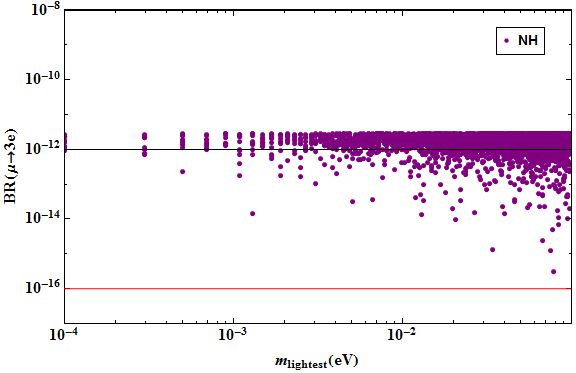}\\
		\includegraphics[width=0.40\textwidth]{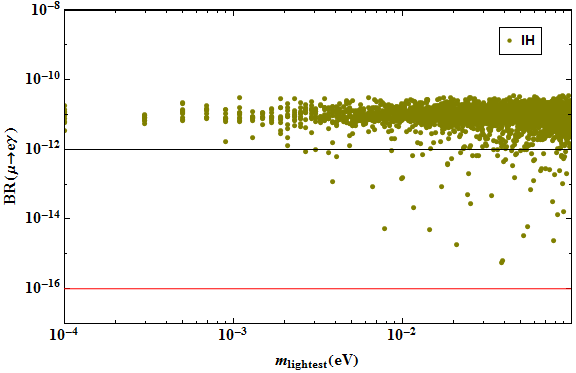}
		\includegraphics[width=0.42\textwidth]{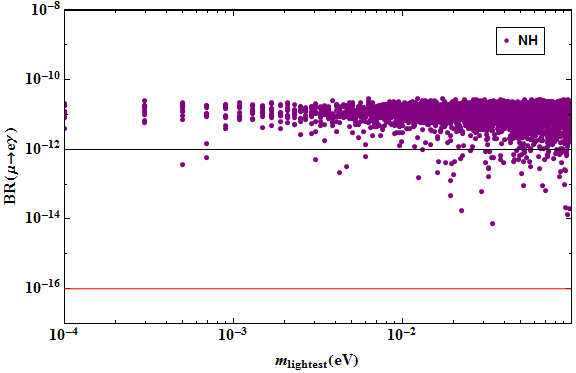}

	\end{center}	
	\begin{center}
		\caption{Total contribution to lepton flavour violation shown as a function of the lightest neutrino mass for in case of type-II dominance case for both  $\mu\rightarrow 3e$ and  $\mu\rightarrow e\gamma$  .The blue and red horizontal line
			shows the limit of BR as given by SINDRUM experiment and the recently proposed limit of $\mu\rightarrow 3e$ experiment
			respectively} 
		\label{fig78}
	\end{center}
\end{figure}

We have correlated lightest neutrino mass and LFV constraints for both type-I and type-II dominant cases considering $\mu\rightarrow e\gamma$ and $\mu\rightarrow 3e$  processes. The BR of these processes has a strong dependency on flavor and heavy neutrino mixing.  $\mu\rightarrow e\gamma$ process dependent on lepton and Higgs coupling whereas $\mu\rightarrow 3e$ is controlled by right-handed neutrino mixing.  We have used the expression
given in \eqref{lfv1} and \eqref{lfv3} to calculate the BR. The lepton Higgs coupling $h_{ij}$  can be computed explicitly for
a given RH neutrino mass matrix by diagonalizing the RH neutrino mass
matrix and obtaining the mixing matrix element, $V_{i}$ and the eigenvalues $M_{i}$. The variation of BR with the lightest neutrino mass for both type-I and type-II dominant cases are shown in the Fig\eqref{fig79} and Fig\eqref{fig78} respectively.

\section{Conclusion}
The quest for NDBD and its interrelation with neutrino mass makes it a very interesting
and an enthralling topic of research at present time.
Neutrino oscillation experiments have already provided us with the first signs of physics
beyond the Standard Model. At the present juncture, the quest for new physics has also got
an unprecedented momentum because of LHC.  Among
neutrino experiments, observation of NDBD would signify lepton number violation and Majorana nature of neutrino mass. However, as is well known NDBD can also occur in many
other scenarios and hence the specific nature of new physics may remain to be ascertained
and in such situations, LHC and LFV processes may provide complementary information.
This interrelation of NDBD with LHC and LFV processes makes it a very engrossing and
interesting topic of research at the present juncture. In this paper, we contemplated the implications of NDBD in the LRSM framework which is realized through $A_{4}\times Z_{2}$ flavor symmetric model, Owing to the presence of new scalars and gauge bosons in this
model, various additional sources would give rise to contributions to NDBD process, which
involves RH neutrinos, RH gauge bosons, scalar Higgs triplets as well as the mixed LH-RH
contributions. We have realized LRSM for both type-I and type-II dominant cases. For a simplified analysis, we have ignored the left-right gauge boson mixing
and heavy light neutrino mixing. We have considered the extra gauge bosons and scalars
to be of the order of TeV. Based on our observations, the following conclusions could be arrived at,

\begin{itemize}
	\item In the standard light neutrino contribution to NDBD, it is observed that, for type-I and type-II dominant case, the effective mass governing NDBD is found to be of the order of $10^{-3}-10^{-1}$ eV in case of the NH and for the IH it is found to be $10^{-2}-10^{-1}$ and are within and much below the current experimental limit. However, in all
	the cases, the light neutrino contribution can saturate the experimental limit for the lightest
	neutrino mass (m1/m3) for (NH/IH) of around 0.1 eV. Variation of effective mass with the lightest neutrino mass is shown in Fig(\ref{fig10})
	\item We have checked the viability of the model by varying  Majorana phase $\alpha$, mixing angle $\theta_{23}$ and CP-violating phase $\delta$ with the effective mass calculated from our model which is found to be well within the experimental limits. These plots are represented in Fig\ref{fig70} and Fig\ref{fig1} for both type-I and type-II dominant cases respectively.
	\item For total contribution considering new physics contribution from heavy neutrino for type-I(NH/IH) dominance case shows results within the recent experimental bound for the lightest mass
	varying from (0.0001-0.1) eV given in Fig(\ref{fig2}).
	
	\item  For total contribution considering new physics contribution from heavy neutrino for type-II(NH/IH) dominance case shows results within the recent experimental bound for lightest mass
	varying from (0.0001-0.1) eV.
	\item  Half-life of the NDBD process, for type-I(NH/IH) and type-II(NH/IH) dominant case light neutrino mass in the range  (0.0001-0.1)eV, and shows results within the experimental bound. Variation of the half-life  of NDBD process with the lightest neutrino mass is shown in Fig(\ref{fig30})
	\item Scalar triplet contribution for type-I(NH/IH) dominance are found to be $10^{-2}$eV and $10^{-3}$eV in the light neutrino mass range (0.0001-0.1)eV. 
	\item Scalar triplet contribution for type-II(NH/IH) dominance are found to be $10^{-3}$eV and $10^{-2}$eV in the light neutrino mass range (0.0001-0.1)eV .
	\item We have also checked the consistency of the model by investigating different LFV processes such as $\mu \rightarrow e\gamma$ and $\mu \rightarrow 3e$. We have shown Variation of branching ratios of these processes with the lightest neutrino mass for both type-I and type-II dominant cases considering NH and IH  in Fig(\ref{fig79}) and Fig(\ref{fig78}) respectively. From these plots, it can be inferred that the type-I dominant case shows results that are more consistent with the experimental bounds.   
\end{itemize}

\section{Acknowledgement}
The research work of MKD and BBB is supported by the Department of Science and Technology, Government of India, under the project grant EMR/2017/001436. BBB would also like to acknowledge Tezpur University institutional grant and Research and Innovation grant DoRD/RIG/10-73/ 1592-A for funding their research work.

\section{Properties of $A_4$ group} 
\label{appen1}

$A_4$ is a discrete group of even permutations of four objects. It has three inequivalent onedimensional representations 1, $1^{\prime}$ and$1^{\prime\prime}$ an irreducible three dimensional representation 3. Product of the singlets and triplets are given by-

\begin{equation}
1 \otimes 1=1 \nonumber
\end{equation} 
\begin{equation}
1^{\prime}\otimes 1^{\prime}=1^{\prime\prime} \nonumber
\end{equation}
\begin{equation}
1^{\prime}\otimes1^{\prime\prime}=1 \nonumber
\end{equation}
\begin{equation}
1^{\prime\prime} \otimes 1^{\prime\prime}=1^{\prime} \nonumber
\end{equation}
\begin{equation}
3\otimes3=1\oplus1^{\prime}\oplus1^{\prime\prime}\oplus 3_{A}\oplus 3_{S}
\end{equation}

where subscripts A and S stands for “asymmetric” and “symmetric” respectively. If we
have two triplets ($a_1$, $a_2$, $a_3$) and ($b_1$, $b_2$, $b_3$), their products are given by

\begin{equation}
1 \approx a_1b_1 + a_2b_3 + a_3b_2 \nonumber
\end{equation}
\begin{equation}
1^\prime \approx a_3b_3 + a_1b_2 + a_2b_1 \nonumber
\end{equation}
\begin{equation}
1^{\prime\prime} \approx a_2b_2 + a_3b_1 + a_1b_3 \nonumber
\end{equation}
\begin{equation}
3_S \approx \left(\begin{array}{c}
2a_{1}b_{1}-a_{2}b_{3}-a_3b_2\\
2a_{3}b_{3}-a_{1}b_{2}-a_2b_1\\
2a_{2}b_{2}-a_{1}b_{3}-a_3b_1 \end{array}\right)\nonumber
\end{equation}
	\begin{equation}
	3_A \approx \left(\begin{array}{c}
	a_{2}b_{3}-a_{3}b_{2}\\
	a_{1}b_{2}-a_2b_1\\
	a_{3}b_{1}-a_{1}b_3 \end{array}\right)\nonumber
	\end{equation}

	\bibliographystyle{utphys}
	\bibliography{bb22}
	\end{document}